\input harvmac
\newcount\figno
\figno=0
\def\fig#1#2#3{
\par\begingroup\parindent=0pt\leftskip=1cm\rightskip=1cm
\parindent=0pt
\baselineskip=11pt
\global\advance\figno by 1
\midinsert
\epsfxsize=#3
\centerline{\epsfbox{#2}}
\vskip 12pt
{\bf Fig. \the\figno:} #1\par
\endinsert\endgroup\par
}
\def\figlabel#1{\xdef#1{\the\figno}}
\def\encadremath#1{\vbox{\hrule\hbox{\vrule\kern8pt\vbox{\kern8pt
\hbox{$\displaystyle #1$}\kern8pt}
\kern8pt\vrule}\hrule}}

\overfullrule=0pt

\def\TD{{T^{(D)}}}
\def\TS{{T^{(S)}}}

\def\TS{{T^{(S)}}}
\def\TD{{T^{(D)}}}
\def\LS{{L^{(S)}}}
\def\LD{{L^{(D)}}}

\Title{MIT-CTP-2566}
{\vbox{\centerline{Non-BPS excitations of D-branes} 
{\centerline{and black holes${}^*$}}}}
\smallskip
\smallskip
\centerline{Samir D. Mathur\foot{E-mail: me@ctpdown.mit.edu}}
\smallskip
\centerline{\it Center for Theoretical Physics}
\centerline{\it Massachussetts Institute of Technology}
\centerline{\it Cambridge, MA 02139, USA}
\bigskip

\medskip

\noindent

This note discusses some results on the non-BPS excitations of D-branes.
We show that the excitation spectrum of a bound state of D-strings
 changes character
 when the length of the wrapping circle
becomes less than $\sim g^{-1}\LS$.  We 
review the observed relation between the low energy absorption cross-section
of D-branes and the low energy absorption cross-section for black holes.
 We discuss
 various issues related to the information question for black holes.

\vskip 1.7in
${}^*$Talk given at Strings '96, Santa Barbara, with
some extensions and additions.
\Date{September 1996}


This is an expanded version of a talk given at
Strings '96, Santa Barbara, with the title `Comparing decay rates
for D-branes and black holes'. Some results on the non-BPS spectra
of higher branes have been extended to cover the case
of 0-branes,  given the interest in 0-branes in this conference.
Some earlier work on the black hole information issue is
reviewed as well.

\newsec{Introduction}

Recently there has been an extensive and fruitful investigation into the
count of the BPS states  in string theory when some chosen
 charges of the configuration are held fixed.
Following suggestions of Susskind \ref\suss{L. Susskind, hep-th/9309145.},
 Russo and Susskind\ref\rs{J. Russo and L. Susskind, {\it Nucl. Phys.}~
{\bf B437} (1995) 611.
} and Vafa\ref\vafa{C. Vafa, \quad\quad{\it unpublished.}}, Sen 
\ref\sen{A. Sen,
{\it Nucl. Phys.}~{\bf B440} (1995) 421 and {\it Mod. Phys. Lett.}
~{\bf A10} (1995) 2081} computed the logarithm of the number  of 
BPS states of the heterotic string with a fixed total charge, 
and found this to equal the area of the  stretched horizon of 
the  corresponding black hole (measured in planck units),
 upto a constant of order unity. The advent of D-branes 
\ref\dbranes{For a recent review and references to original
literature see J. Polchinski, S. Chaudhuri and C.V. Johnson,
 ``{\it
Notes on D-branes}'', hep-th/9602052.} provided objects in
 string theory carrying a variety of different charges. 
Strominger and Vafa \ref\stromvafa{A. Strominger and C. Vafa,
hep-th/9601029} constructed  a black hole model in 4+1 
dimensions carrying three different charges and thus
 possessing a nonsingular horizon. They found that the 
entropy for branes with a given set of  charges {\it exactly}
 equalled the Bekenstein-Hawking entropy for a black hole 
carrying the same charges. Other examples of this 
correspondence were soon developed \ref\horo{For a review 
of black hole entropy in string
theory and references, see G. Horowitz, gr-qc/9604051}\ref\kleb{I.R.
 Klebanov and A.A. Tsetlin, hepth 9604166.}.  These results suggest 
that the 
Bekenstein entropy defined by
the area of the horizon is in some way a count of the possible 
microstates of the black hole, 
though it is not yet clear where these microstates reside. They 
also provides a striking validation of string theory, with the 
large number of perturbative and non-perturbative particle
 species in the theory finding a natural place in reproducing
 the Bekenstein-Hawking entropy predicted by  classical gravity and
quantum field theory.

To investigate issues of Hawking radiation and information loss, 
we must consider non-BPS
states of the theory, since extremal holes do not radiate. 
When investigating the physics in this domain one has
to be more careful, since we do not have the non-renormalisation 
theorems that applied to the case of BPS-states.
But there are several results \ref\callanmalda{C.G. Callan and
 J. Maldacena,
hepth/9602043.}\ref\many{G. Horowitz and A. Strominger, 
hep-th/9602051; G. Horowitz,
 J. Maldacena and
A. Strominger, hep-th/9603109; S.S. Gubser, I. Klebanov
and A.W. Peet, hep-th/9602135; I. Klebanov and A. Tseytlin,
hep-th/9604089;M.
Cvetic and D. Youm, hepth/9603147, hep-th/9605051, hep-th/9606033;
G. Horowitz, D. Lowe and J. Maldacena, hep-th/9603195; M.Cvetic
 and
A. Tseytlin, hep-th/9606033;
E. Halyo, A. Rajaraman and L. Susskind, hep-th 9605112.}\ref\spenta{A. 
Dhar, G. Mandal and
 S.R. Wadia, hep-th/9605234.}\ref\dasmathurtwo{S.R. Das and S.D. Mathur,
hep-th/9606185}\ref\dasmathurthree{S.R. Das and S.D. Mathur,
hep-th/9607149} that encourage the belief that the physics 
captured by
the regime of non-BPS D-brane physics where we are able to do 
computations, 
is in some way related to the physics of black holes, at
least at low energies.

In this note we do the following:

(a)\quad We examine the spectrum of a bound state of 0-branes, when
the spacetime has been compactified on a circle. The spectrum exhibits 
some
curious features when the scale of compactification becomes smaller
than the natural size of the 0-brane bound state. The spectrum in this
 domain is
found by using dualities on the known
 spectrum of the elementary string, following the methods in
 \dasmathurthree .

(b) \quad We turn the above process around, starting from the spectrum
of the bound state of $n_w$ 0-branes when the
compactification scale is large (essentially noncompact spacetime)
 and obtaining the excitations for the bound state of $n_w$ D-strings
when the 
length of the wrapping circle is smaller than $L_{\rm cr}\equiv g^{-1}\LS$. 
(Here $g=e^\phi$
is the elementary string coupling and $\LS=(\alpha')^{1/2}$
is the length scale of the elementary string.) We thus find that the
lowest lying excitations for the D-string change character as the length
of the wrapping circle $L$ 
drops below $L_{\rm cr}$: for $L>> L_{\rm cr}$ we have the 
universal spectrum of vibrations of a single string of length $n_w L$,
while for $L< L_{\rm cr}$ we get position independent 
oscillations coming from the nonzero thickness of
the  bound state.

(c)\quad We review the computation for the absorption cross section
of low energy quanta into a combination of branes. The branes
are chosen to carry the charges of a black hole with  nonzero horizon
area in 4+1 dimensions, following \stromvafa\callanmalda . 
The absorption cross section
of the quanta studied agrees with the low energy absorption
cross section for the corresponding black hole. 

(d)\quad We discuss the black hole information paradox. In 
particular we discuss
the large quantum gravity effects that appear
to exist in some approaches \ref\SVV{E.~Verlinde and H.~Verlinde,
 hep-th 9302022;
 K.~Schoutens, E.~Verlinde and H.~Verlinde,
{\sl Phys. Rev.} {\bf D48} (1993) 2670; Y.~Kiem, E.~Verlinde 
and H.~Verlinde,
{\sl Phys. Rev.} {\bf D52} (1995) 7053.
}\ref\klmo{E.~Keski-Vakkuri, G.~Lifschytz,
 S.~D.~Mathur and
               M.~E.~Ortiz, {\sl Phys. Rev.} {\bf D 51} (1995) 1764.} 
to perform the full calculation of Hawking radiation, 
and argue that these are only an apparent effect arising from a 
turning point in
 a semiclassical
quantum gravity wavefunction \ref\eskomathur{E. Keski Vakkuri and 
S.D. Mathur, gr-qc 9604058 (To appear in Phys. Rev. D)}. We
 review the breakdown of the classical field limit in 1+1 
dimensional string theory  before the threshold of black hole formation \ref\dasmathurzero{S.R. Das and S.D. Mathur, Phys. Lett. {\bf B365} 
(1996) 79, (hep-th 9507141).}.  We discuss the possible implications of 
the
results on the cross
section agreements between D-branes and black holes, and the
issue of measurability of black hole hair.

\newsec{Non-BPS spectrum of 0-branes and 1-branes.}

Let the spacetime be flat Minkowski $M^9\times S^1$, where the
coordinate $X^9$ has been compactified on
a circle. Let $g=e^\phi$ be the elementary string coupling.
Let $\LS$ be the length scale associated with the elementary string:
\eqn\five{\LS=(2\pi)^{1/2}(\TS)^{-1/2}}
and $\LD$ be the length scale associated with the  D-string of type
 IIB string
theory:
\eqn\six{\LD=(2\pi)^{1/2}(\TD)^{-1/2}, ~~\TD=\TS/g}
Let the D-string have winding number $n_w$ around $X^9$, and no momentum
along $X^9$. Let the length of the compactified circle be
\eqn\seven{L=A\LD=A\LS g^{1/2}}
If we T-dualise in the
compact direction $X^9$, the new length of the circle will be
\eqn\eight{L'=A^{-1}g^{-1/2}\LS}
and the new coupling will be
\eqn\nine{g'=g[L'/L]^{1/2}=g[A^{-2}g^{-1}]^{1/2}=g^{1/2}A^{-1}}
The D-string of type IIB string theory 
will change to a bound state of $n_w$ 0-branes
of type IIA theory, with no momentum in the $X^9$ direction.

{\subsec {Spectra}}

We know the following about the spectrum of the elementary string. Suppose 
the radius of the $X^9$ circle is
\eqn\ten{L_1=A\LS}
Let the elementary string have winding number
$n_w$ around $X^9$, and no momentum along $X^9$. If $g=e^\phi<<1$, and $A>1$, 
the we have a a spectrum of long lived
excitations for the low lying states, given by essentially the
free string spectrum.  The 9-dimensional masses of the string states are
\eqn\eleven{m^2=(n_w L_1\TS)^2+8\pi\TS N}
where $N$ is the excitation level over the 
ground state in both the right and left sectors
(which can each be either  Ramond  (R) or Neveu-Schwarz (NS)).
 The term `long lived excitation' used above 
stands for the fact that the lifetime of the excited state  
with excitation energy $\Delta E$ is much larger than $(\Delta E)^{-1}$.  
The restriction on $A$ may be relaxed somewhat, 
but we cannot let $A$ become too small for nonzero $g$, 
for then the spectrum changes, as we will discuss later. 
The restriction $A>1$ is a convenient starting point for our present 
purposes.
 The reason for
the change of spectrum is that if there is a very small compactified
circle, then there is a very low 
mass winding state that can contribute in loop corrections to the 
string eigenstate. 

By S-duality of the type IIB theory, we can conclude the following for 
the spectrum of the D-string. Suppose we have $g_D=g^{-1}<<1$, 
and the length of the $X^9$ circle is
\eqn\elevenp{L=A\LD=A\LS g^{1/2}}
and $A>1$. Then we have a spectrum of low lying long lived
excitations given by
\eqn\twelve{m^2=(n_w L\TD)^2+8\pi\TD N}
For $n_w$, $A$, fixed, and $N>>n_w^2A^2$, we get
\eqn\thirteen{m\approx (\TD)^{1/2}\sqrt{8\pi}\sqrt{N}}
For $n_w$, $N$ fixed, and $A>>n_w^{-1}\sqrt{N}$, we get
\eqn\fourteen{m\approx (\TD)^{1/2}\sqrt{2\pi}n_w A +(\TD)^{1/2}
{2\sqrt{2\pi}   \over n_w A }N}

On T-dualisation we get the 0-brane state with the same energy levels.
  Expressed in terms of $g'$ rather than $g$ the energy levels are for
 $n_w$, 
$A$, fixed and $N>>n_w^2A^2$:
\eqn\fifteen{m\approx (\TS)^{1/2}\sqrt{8\pi}(g')^{-1}A^{-1}\sqrt{N}}
and for $n_w$, $N$ fixed and $A>>n_w^{-1}\sqrt{N}$:
\eqn\sixteen{m\approx \TD_0 n_w +
(\TS)^{1/2}(g')^{-1}{2\sqrt{2\pi}   \over n_w A^2 }N}
where
\eqn\seventeen{\TD_0=\TS(g')^{-1}\LS=(\TS)^{1/2}(g')^{-1}\sqrt{2\pi}}
is the tension (i.e. the mass) of the 0-brane.

Now we discuss what we know directly from 
the spectrum of bound states for the 0-branes. Let
\eqn\eighteen{L_0\equiv (g')^{1/3}\LS}
In \ref\danielsson{U.H. Danielsson, G. Ferreti and  B. Sundborg, hepth 9603081.}\ref\kabat{D. Kabat and P. Pouliot, 
hepth 9603127.}\ref\douglasetal{M.R. Douglas, D. Kabat, 
P. Pouliot and S. Shenker, hepth 9608024.} it was argued that for 
\eqn\nineteen{g'<<1~~~~L'>>L_0,}
the length scale of the
bound state of $n_w>1$ zero branes is $\sim L_0$. Further 
it was
argued that the low lying non-BPS excitations (carrying no net charge) 
are
given by levels with spacing of order
\eqn\twenty{(\Delta E)_0\sim (g')^{1/3}(\TS)^{1/2}}
It is not immediately clear, however, that there is any 
discernable level structure in the low level excitations of the 
zero brane
 bound state.
The argument of \danielsson\  used a separation of slow and
fast modes to do a semiclassical expansion; there is however no
small parameter that actually governs this separation. There may
 exist broad
resonances with width of the same order as the height,
exhibiting structure at the scale \twenty , but there is no clear 
evidence for
this either. It is true, however, on dimensional grounds, that 
the only scale
exhibited by the excitations is that given by \twenty .

For later use we note that
\eqn\twentyone{L'/L_0=A^{-2}(g')^{-4/3}=A^{-2/3}g^{-2/3}.}

{\subsec {Elementary string $\rightarrow$ D-string $\rightarrow$ 0-branes}}

The spectrum \twelve\  for the D-string was obtained for $g>>1$, $A>1$. 
From \eight\ we have that $L'<<\LS$. 
Using \nine\ we see that  we can choose $A\sim 1$ to get
$g'>>1$, or choose $A$ sufficiently large so that we have $g'<<1$. 
Let us make the latter choice. Then since the spectrum does
not alter under T-duality, we find from \sixteen\ that for the type IIA theory
0-brane bound state we have long lived excitations with separation
\eqn\twentytwo{(\Delta E)_1={2L'\over n_w}\TS}

The spectrum \twentytwo\ is very different from \twenty .
 In obtaining \twentytwo\
 we
have used a parameter range where
\eqn\twentythree{L'/L_0=A^{-2/3}g^{-2/3}<<1}
where the  inequality follows because $A>1$ and $g>>1$. Thus
the 0-brane state has been `squashed' in the compact direction to a size
much smaller than the natural scale of 0-brane bound states. 

We can use this result to speculate on the structure of  0-brane
bound  states.
The scale 
 of excitations \twenty\  can be understood heuristically in the 
following way.
For a bound state  of two 0-branes, say, we get an excited state by
attaching a pair of open strings
with the opposite orientation, beginning at one 0-brane and ending at 
another.
 If the 0-branes are a distance
$l$ apart, then the energy from the tension of these strings is
$V\sim l\TS$. On the other hand confining the 0-branes within a region
of size
$\sim l$ gives a kinetic energy for each 0-brane 
$K\approx p^2/(2M)\sim l^{-2}
(\TS)^{-1/2}g'$ ($M$ is the mass of the 0-brane).
 Minimising the total energy
of excitation
$(\Delta E)_0=V+T$ gives $l\sim (\TS)^{-1/2}(g')^{1/3}$,
 and $(\Delta E)_0\sim
(\TS)^{1/2}(g')^{1/3}$, in accordance with \twenty .

If we were doing the quantum mechanics of two pointlike
objects, after compactification of $X^9$ to a small circle the argument
of the preceeding paragraph would still apply, and again yield
the scale \twenty . The wavefunctions would simply reduce
to constants in the compact direction. But if the average separation 
between
the 0-branes is $\sim L_0$, then the open strings stretching from one
0-brane to the other would give an energy scale $\sim\L_0\TS$ which is much
larger than \twenty . If we start and end the open strings on the same
 zero brane,
while wrapping it on the compact circle $X^9$, then we get the energy levels
\eqn\twentyfour{(\Delta E)_2=2NL'\TS}
which differs from \twentytwo\  by the factor $n_w$. 

What physical picture can give the extra $n_w$ in \twentytwo ? Since
we have kept $g'<<1$, it is tempting to look for a picture of the excitation
in terms of a small number of open strings, though this might
be invalid due to loop corrections in the presence 
of the very small compactification scale. We list
three possibilities:

(1) \quad We must use fractional open strings, with tension $(n_w)^{-1}\TS$ 
in \twentyfour , following the notion of fractional branes discussed in \ref\maldasusskind{J. Maldacena 
and
L. Susskind, hepth/9604042.}.

(2) \quad The natural scale of the 0-brane bound state is $\sim L_0$
in noncompact space, but if one direction is compactified
to a length much smaller than $L_0$ then the bound state becomes reduced
to that compactification scale in all directions.  If this happens then the
open strings stretching from one 0-brane to another may yield  levels
of the order \twentytwo , though there is no immediate reason
 for this precise form.

(3) \quad The 0-branes in the compactified spacetime have a disclike shape,
(with perhaps the scale $L_0$ in the noncompact direction), and these discs
are stacked parallel to each other with separations $L'/n_w$ along the compact
direction $X^9$. The open strings can stretch from one disc to the next one,
starting at any point on the first disc and stretched parallel
 to the direction $X^9$. The wavefunction of the
open string is a uniform superposition over the various possible locations
of the  end point on the first disc. (In fact to get a correct count of BPS
 states it appears more natural to use open strings with fractional
 tension here just as in (1) above, with the minimum excitation 
involving fractional open strings that stretch from each 0-brane to the next.)

If possibility (3) is correct then it could be interesting
for the following reason. D-brane excitations  at weak coupling
(and no small compactified directions) are given 
by open strings that are attcahed to a hyperplanes that are infinitely
thin in the Dirichlet directions. But the ideas of Susskind about
 black holes suggest that at strong coupling the D-branes
should be described by an effective theory that has open strings ending
on an extended surface (the horizon) which is not itself the surfaces of
the D-branes 
that the black hole was constructed with. The description (3) above
also requires an effective extended endpoint for the open 
strings attached to a
 0-brane.

{\subsec {0-branes $\rightarrow$ D-strings.}}

 Let us now start from the other side, with a bound state
of $n_w>1$ 0-branes, in a domain of parameters where we know something
about the spectrum: 
\eqn\twentyfive{g'<<1,~~~~L'/L_0>>1}
 Then as mentioned above,
the spectrum has structure at the energy scale \twenty , and this will
 be also the
structure of the spectrum of any string or brane obtained through dualities.
From \twentyone , we have
\eqn\twentysix{A<<(g)^{-1}}
\eqn\twentyseven{g=(g')^2A^2<<(g')^{2/3}<<1 }
\eqn\twentyeight{L=A\LD<<g^{-1}\LD=\LD g_D=g^{-1/2}\LS}
Thus we have weak elementary string coupling $g$ and
the length $L$ of the D-string much longer than the
elementary string scale $\LS$. At first we might expect that in
this situation we would get the spectrum of excitations given by
attaching open strings to the D-strings. If the $n_w$ D-string bound state
implies just the naive 
$n_w$ valued Chan-Paton factors at the ends of the open strings 
 then the spectrum would be (for no net momentum in the $X^9$ direction)
\eqn\twentynine{E_N={4\pi N\over L}={2\sqrt{2\pi} N\over A}g^{-1/2}\TS^{1/2}}
with degeneracy $n_w^2$ for each level. If the $n_w$ D-strings behave as one
string of length $n_w L$ then the spectrum would be
\eqn\thirty{E_N={4\pi N\over n_w L}={2\sqrt{2\pi} N\over n_w A}g^{-1/2}\TS^{1/2}}
with degeneracy unity for each level. 

What we actually have from \twenty\ by duality is structure in 
the energy spectrum at
the scale
\eqn\thirtyone{(\Delta E)_0\sim g^{1/6}A^{-1/3}\TS^{1/2}}
From \twentysix , we see that the range of validity of our
 analysis is $A<g^{-1}$. 
For
$A\sim g^{-1}$, $n_w$ small, 
 all the scales \twentynine , \thirty , \thirtyone\ 
 are $\sim g^{1/2}\TS^{1/2}$. But as we reduce
$A$ below $\sim g^{-1}$, the levels \twentynine , \thirty\ 
 become higher than the scale \thirtyone\  at which we first
see the structure of  excitations. 

The above observation may be relevant to the consideration
of non-BPS entropy and absorption coefficients of D-brane configurations
 that
are anticipated to resemble black holes. The low energy spectrum used
in \callanmalda\ was analogous to \twentynine , and that used
 in \maldasusskind\ was analogous to \thirty .
But as we reduce the length of the D-string, which happens as we 
reduce the compactification
scale to go towards a black hole,  we find that  excitations at the
 scale \thirtyone\ 
dominate the low energy physics.

At an intuitive level, the appearance of the scale \thirtyone\  might be
understood as follows. If the D-string is very long (longer than
$\sim g^{-1}\LS$)  then the thickness of the strands
making up the string is less than the typical separation between the
strands in the process of oscillation. Thus we simply get the universal
spectrum {} of one string of length $n_w L$. But for the D-string shorter than
$\sim g^{-1}\LS$ the `breather modes' of the thick
soliton are of lower energy than the 
universal oscillation modes of the string, and  dominate the low energy
excitations. The timescale of these latter oscillations are probably 
the same as the timescale for dissociation of the bound state, so 
it is not clear if these should be thought of as oscillations at all.

Using S-duality we can  state the result corresponding to  \thirtyone\ 
for the elementary string.
If we take an elementary string at large coupling $g>>1$,  
wound on a circle of length smaller than $g\LS $, then we will
get an excitation spectrum that has structure  at scale
\eqn\thirtytwo{\Delta E \sim g^{-2/3} A^{-1/3}(\TS)^{1/2}}

{\newsec {Absorption into D-branes}}

We consider the absorption of low energy quanta into extremal black holes
in 4+1 spacetime dimensions, and compare this to the absorption by 
D-branes carrying the same charges as the black hole 
\dasmathurtwo \dasmathurthree . Let the spacetime be
$M^6\times T^5$, where the directions
$X^5\dots X^9$ have been compactified on the torus $T^5$. The black hole 
must carry three nonzero charges in order to have a classically
 nonvanishing area of the horizon. Following 
\stromvafa \callanmalda\ we make the corresponding 
D-brane configuration by taking one D-5-brane wrapped
on the torus, a D-string wrapped $n_w$ times around the $X^9$ 
cycle and bound to this D-5-brane, and let the D-string carry
 momentum along the $X^9$ direction. 

A D-string of length $L$ may be considered as a system with some discrete
energy levels with spacing $\Delta E$ which is independent of $E$. 
Consider an initial state at $t = 0$
where the D-brane system is in its BPS
ground state and a massless closed string state of energy $k_0$ is 
incident on it.
Let the amplitude to excite the D-string to  any one of the excited levels
per unit time be $R$.
(For $t$ large, only the levels in a narrow band will contribute,
and in this band we can use the same $R$ for each level.)
 Then the amplitude that the system in an
excited state with energy $E_n$ at a given time $t$ is given by
\eqn\tenone{ A(t) = R e^{-iE_nt}\int_0^t dt' e^{i(E_n - k_0)t'}
= Re^{-{i\over 2}(E_n+k_0)t}[{2 \sin [(E_n-k_0)t/2] \over (E_n-k_0)}]}
The total number of quanta absorbed in time $t$ is thus given by 
\eqn\tentwo{P(t) = \sum_n |R|^2 [{2 \sin [(E_n-k_0)t/2 ]\over (E_n-k_0)}]^2
\rho (k_0)}
where $\rho (k_0)$ denotes the occupation density of state of the
 incoming quantum. For large length of the D-string $L$ we can replace 
the sum by 
an integral
\eqn\tenthree{\sum_n \rightarrow \int {dE \over \Delta E}}
in which case the rate of absorption $ {\cal R_A} = P(t)/t$ evaluates to
\eqn\tenfour{{\cal  R_A }(t) = {2\pi |R|^2 \over \Delta E} \rho( k_0)}

For our case of the D-string on the 5-brane, 
\eqn\aone{\Delta E~=~{4\pi\over n_w L} }

Here we have used the fact that for a sufficiently large 
wrapping radius or sufficiently large $g$ a bound 
state of D-strings exhibits the excitation spectrum 
of a single multi-wound string of length $n_w L$ 
\ref\dasmathurone{S.R. Das and S.D. Mathur, hepth 9601152}.

Consider the absorption of a quantum of the 10-dimensional 
graviton $h_{12}$, with no
momentum or winding along the compact directions. 
This is a neutral massless scalar of the 5-dimensional
 theory. There are two open
string states that can be created on the D-string in absorbing this
graviton. We can have the string with polarisation $1$ travelling
left on the D-string and the open string with polarisation $2$
travelling right, or we can have the polarisations the other way
round. This means that there are two series of closely spaced levels
that will do the absorption, and so the final rate of absorption
computed from \tenfour\ will have to be doubled.

 To find $R$, we have to examine the action for the D-string coupled to 
gravity. Writing the action with only the
fields that we will use below:
\eqn\qone{S={1\over 2\kappa^2}\int d^{10}X[R-{1\over 2}(\partial\phi)^2]+
T\int d^2\xi e^{-\phi/2}\sqrt{\det[G_{mn}]} }
where
\eqn\qtwo{G_{mn}=G_{\mu\nu}(X)\partial_m X^\mu\partial_n X^\nu}
and $T$ is a tension related to the tension of the elementary string by
$\TS=e^{\phi/2}T$. Note that the tension of the D-string is
$\TD=Te^{-\phi/2}$. Expanding this action to lowest required order,
with $G_{\mu\nu}=\eta_{\mu\nu}+2\kappa h_{\mu\nu}$:

\eqn\qthree{S\rightarrow \int d^{10}X {1\over 2}(\partial h_{ij})
(\partial h^{ij})+{1\over 2}(\delta_{ij}+2\kappa h_{ij})
\partial_\alpha (\sqrt{\TD}X)^i\partial^\alpha (\sqrt{\TD}X)^j }

From \qthree\ we find for the amplitude per unit time for the graviton
to create any one of these two possible open string configurations to
be
\eqn\atwo{R=\sqrt{2} \kappa |p_1| {1\over \sqrt{2k^0 }}
{1\over \sqrt{L}} {1\over \sqrt{V_c}} {1\over \sqrt{V_T}} 
\rho_L^{(1/2)}(|p_1|) }
$p_1$ is the momentum of the massless open string travelling left, say,
while $k^0$ is the energy of the absorbed quantum.
Here we have separated the term ${1\over \sqrt{V}}$ into
contributions from the string direction $X^9$, the remaining four
compact directions (denoted by the subscript $c$) and the transverse
noncompact spatial directions (denoted by the subscript $T$). We have
also included the term
\eqn\afour{[\rho_L(|p_1|)]^{1/2}=[{T_L\over |p_1|}]^{(1/2)}, ~~~
 T_L={S_L\over \pi n_w L} } 
which gives the Bose enhancement factor
 due to the population of left moving open string states on the
 D-string \callanmalda .  Here $S_L$ is the entropy of the extremal
 configuration, given by the count of the possible ways to distribute
 the $N$ quanta of momentum among different left moving vibrations of
 the D-string:
\eqn\eone{S_L~=~2\pi\sqrt{n_w N} }
and equals the Bekenstein entropy of the black hole with the same
 charges as the D-brane configuration.

The absorption cross section is given by
\eqn\athree{\sigma~=~2{\cal R_A}/{\cal F}}
where ${\cal F}=\rho(k_0)V_T^{-1}$ is the flux, and the factor
of $2$ was explained before eq. \atwo. 

Note that
\eqn\afive{{\kappa^2\over LV_c}~=~8\pi G_N^5}
and that for the given choice of momenta
\eqn\gone{k_0~=~2|p_1|  }
Then we find
\eqn\asix{\sigma~=~A}
where $A=8\pi G_N^5\sqrt{n_w N}$ is the area of the extremal black
hole with one 5-D-brane, $n_w$ windings of the 1-D-brane, and momentum
charge $N$. 

To compare this to the classical absorption cross section 
at low energies, one solves the wave equation for the 
incoming quantum in the metric of the black hole 
(using the approximation that the wavelength is
 much larger than the Schwarzschild radius of the black hole). 
Such a calculation for the 4+1 dimensionsal hole is performed in
 \dasmathurtwo , following 
the calculation for 3+1 dimensions in \ref\unruh{ W.G. Unruh, {\it
 Phys. Rev.}~{\bf D14} (1976) 3251.}. The result is precisely \asix . 
Thus we get agreement between the absorption cross-sections of the 
D-branes on the one hand, and the black hole they will form for  
a different choice of coupling on the other.

In \dasmathurthree\ it was shown that the cross section for
 the dilaton agrees as well between the black hole and 
the D-brane configuration.
Note that the D-string can only oscillate within the 5-D-brane, 
which is wrapped on the  internal directions. Thus at this order of 
calculation, only 5-dimensional scalars are absorbed, and vectors 
and gravitons have no absorption cross-section. This agrees 
with the fact that for 3+1 dimensional holes the low energy 
cross section vanishes for vectors and gravitons 
\ref\staro{ A.A. Starobinsky and S.M. Churilov,
Sov. Phys.- JETP {\bf 38}, 1 (1974)
}. By supersymmetry, we expect that the classical cross section 
for spin 1/2 quanta is 
related to that for scalars, and is thus $\sim A$ as well. 
The D-brane calculation yields the  order $\sim A$ also, 
since we can have a fermionic open string as the right mover
 and the bosonic open string again as the left mover, 
giving a spin-1/2 quantum in the emitted state.  

In the above calculation we observe that the cross section for 
low energy massless neutral scalars was precisely the horizon 
area, in 5 dimensions. This is also the case in 4 dimensions,
 so one wonders 
if for such quanta one always gets the area of the horizon,  
 and if there are further universalities
among the low energy cross sections for particles with various spin.
 This issue is addressed in \ref\dmg{S.R. Das, G. Gibbons and S.D. Mathur, 
{\it in preparation}.}, where it is shown that such is indeed the case.

Recently there has been interesting progress on this issue.
The above calculation for neutral scalars has been extended to
 charged quanta in both 4 and 5 dimensions 
\ref\klebgub{S. Gubser and I.R. Klebanov, hepth 9608108. }
Further, it was shown in \ref\maldastrom{J. Maldacena and A. Strominger, 
hepth 9609026.} that  the D-brane configuration reproduces 
the characteristics of black hole grey-body factors 
for both neutral and charged quanta. 

{\newsec {Black hole information}}

{\subsec {Quantum gravity effects}}

It has been often felt that the calculation of Hawking that gives
Hawking radiation in an essentially thermal form should suffer from
quantum gravity corrections that might permit information to leak out
 with the 
radiation. This possibility has been stressed by 't Hooft, and a calculation
was performed in \SVV\ which indicated that commutators of operators
associated to the radiation
in a full theory of quantum gravity became large at the horizon. 
Similar large effects were found in \klmo\ in a 
Hamiltonian formalism.

Should we consider this as convincing evidence that the information
paradox is in fact just an effect of using semiclassical gravity where
quantum gravity had to be used? In \eskomathur\ it was argued
that such need not be the case. In investigating the issue of
unitarity, it is best to work with states rather than operators, since
if one uses operators there still remains the question of which states
the operators must be sandwiched between to decide if large quantum
gravity effects exist. 

In the Hamiltonian formulation of quantum gravity (as opposed to
the path-integral formulation) the wavefunction is a function only
of spacelike 3-geometries and matter on the 3-geometries (for the case of
3+1 dimensionsional spacetime physics). The time direction arises 
as the phase of the wavefunction of the 3-geometries, in a WKB
 approximation of this wavefunction \ref\banks{V.~G.~Lapchinsky 
and V.~A.~Rubakov,
 {\sl Acta Phys. Pol.}
             {\bf B10} (1979) 1041; T.~Banks, {\sl Nucl. Phys.} {\bf B249}
             (1985) 332.
}. If the matter density is
small compared to planck density, we have a Born-Oppenheimer approximation
where the `fast' variable is the metric  and the `slow' variable is
the matter (i.e. the phase of the variable representing gravity
oscillates much faster than the phase of the matter variable). The
quantum matter obeys a Schrodinger equation in the `time' that
emerges in this WKB approximation.

But if spacetime is obtained in this way, then we have the issue of
what happens when the fast variable (the parameter labelling the
3-geometries) encounters a turning point. At a turning point the fast
variable would not be fast any more, and we can expect that the
semiclassical description will be invalid. As the gravity variable
recedes from the turning point and becomes `fast' again, the
semiclassical gravity description is valid again, but we can ask for
the total `damage' created by the turning point: namely the error created
by using the semiclassical description through the turning point where
it was not really valid.

This `damage' was computed in several simpe examples in \eskomathur , and
it was found that while the semiclassical description indeed breaks down 
in the vicinity of the turning point, the departure from semiclassicality
erases itself as we recede from the turning point, at least in simple models
of quantum gravity, leaving a small but computable net effect of
the temporary loss of semiclassicality.   This `miraculous' cancellation
of large departures from semiclassicality can be traced to the
fact that by a canonical transformation a turning point can be relocated or
removed.

In the black hole context the
large temporary departures from semiclassicality appear as large
 quantum gravity 
effects in the Schwarzschild coordinate near the horizon. It is plausible that
 these large effects are closely related to the large commutator
 effects seen in \SVV . 
Thus
we conclude that large quantum gravity effects may be only apparent
effects and not real effects, at least in simple models of quantum
 gravity which 
do not involve
 extended objects like strings or branes.

{\subsec {1+1 dimensional noncritical string theory}}

String theory provides a renormalisable model of quantum gravity, 
so one would like to investigate black hole information issues 
using string theory. Computing just a few orders of perturbation theory, 
however,
 is unlikely to yield insights on this problem, beyond what is known from
 semiclassical gravity. 
Luckily it is possible to perform a sum over all string loop diagrams 
in the
case of the noncritical 1+1 dimensional string. As shown in
 \ref\dasjevicki{S.R. Das and A. Jevicki, Mod. 
Phys. Lett {\bf A5} (1990) 1639.}
the $c=1$ random surface model can be recast as a theory of 
free fermions,
where the fermi surface profile acts like the gradient of 
a scalar field defined
over a 1-dimensional spacelike dimension, with time being 
the second dimension.

As observed by Polchinski \ref\polchinski{J. Polchinski, Nucl. Phys.
 {\bf B362} (1991) 125}, it is possible for the fermi surface to
form `folds' in the process of evolution, which destroys the immediate
relation between the local fermi level and the value of the scalar field.
In particular this effect occurs before the threshold of `black hole
formation' in this theory. We can take a coherent pulse made from the quanta 
 of the scalar field, and have this pulse move in from infinity towards the 
strong coupling `wall' present in the this model in the vacuum. For low 
amplitudes of the incident pulse, a slightly distorted but still coherent
 pulse returns after reflection from the wall. But as the amplitude of the 
initial pulse exceeds a certain threshold, the returning pulse develops 
a `fold' in the fermi surface, in the description through free fermions.
 What does this mean in terms of the scalar field description? 

In \dasmathurzero\ it was shown that the scalar field state
 after fold formation corresponds to a collection of incoherent quanta,
 with frequencies that range from low values to very high values: the
 average frequancy amplification over the frequency of the initial pulse 
is of the order of the square root of the number of quanta in the initial
 state. In particular this is not the profile expected of thermal
 radiation that may result from a process of virtual black hole 
formation and evaporation. 
It appears that as the initial pulse enters a strong coupling area,
 the approximation that the string theory is a theory of a single 
particle species (the tachyon) breaks down, and stringy effects
create an outgoing state of a form that cannot arise in a field 
theory with just the tachyon field. It would be interesting if
this phenomenon were to happen in higher dimension string theories
 as well,  as we approach the threshold of black hole formation.

{\subsec {Scattering off black holes}}

At least for extremal and near extremal black holes, there is now
a fairly convincing case that that the Bekenstein entropy
should be interpreted as some count of possible microstates.
Can we scatter quanta off a black hole, and in the process 
determine which state
the black hole is in? This would indicate that black holes
 are not really `black', 
and are much like ordinary particles. We have seen 
 in the last section that
the absorption cross section of D-branes matches that
 of black holes
at low energies. But with the D-branes, if we do
 scattering experiments we
would indeed know which microstate the branes were in.
 For example, suppose
we examine the absorption of a 5 dimensional scalar of energy
 ${4\pi N\over n_w L}$, in
the notation of the last section. Then the absorption of this scalar creates
a pair of
open strings on the D-string with energy ${2\pi N\over n_w L}$ each, and
a certain polarisation. The absorption probablility is directly proportional to
the number of open strings that already inhabit the  state of the
left moving open string,through the
 Bose enhancement factor. But all the microstates that give the
 same mass and charges to the D-brane configuration do not have 
the same number of
quanta inhabiting this particular open string state, and so they have
different absorption
cross sections for the chosen incoming quantum. Thus by
patient experimentation, information on the actual microstate can
be deduced from absorption/scattering processes.

Since the black holes that correspond (in a different 
range of parameters)
to this D-brane cluster have the same formula for the low 
energy cross section
one may think that we can also find the microstate of 
a black hole by the same
process. But here we make some observations that indicate that we have
to be more careful before reaching such a conclusion.

Let us fix the charges and consider different ranges of $g$. 
For $g$ too small,
we presumably get  thick solitonic strings and branes, much thicker 
than the  Schwarzschild radius of the configuration. The case 
of $g$ very large is just
the S-dual of the elementary string with $g$ small, so that 
we get solitonic  5-branes with an elementary string inside. 
This does not look like the black hole we want.  
Let us therefore take $g\sim 1$ in the discussion below.

In the D-brane calculation presented in the
last section the absorbing levels were discrete, long
lived levels. This was the case because we took a D-string that was
wrapped on a very long circle, much longer than $g^{-1}\LS$. But if
we wish to go towards the black hole limit of D-branes, we need
to take a scale of compactifcation that is not too large for a given
number of branes.

 But as we reduce this length of the wrapping circle 
of the D-string below  $g^{-1}\LS$, we saw in section 2 that the spectrum
of excitations at low energies changes to one that has no
sharp levels, just broad resonances at best, with the width of
 the resonances 
comparable to the height. The latter circumstance just means
that the lifetime of the excited state is comparable to the time-scale
$(\Delta E)^{-1}$, where $\Delta E$ is the typical separation between
levels. This feature was a reflection of the fact that there is no
scale in the 0-brane bound state spectrum (to which the D-string spectrum 
is dual when the compactification length is smaller than $g^{-1}\LS$)
apart from the overall energy scale $(g')^{1/3}\TS^{1/2}$. 

Note that the above discussion is for a D-string in isolation 
and not inside a 5-D-brane. But let us assume that the above
 change of spectrum persists in the latter case as well.
If we lose the picture of discrete levels before we reach the black
hole limit of the D-brane configuraton, then it is not immediately clear
what features can be picked up about black holes 
in scattering experiments. It is still
true that with enough diligence we can extract all information
from the D-brane system which is not in the black hole limit. But
 we see that 
the encoding of this information in the results of scattering
 experiments begins to change as we approach the black hole limit.

{\subsec {Low energy absorption cross-sections}}

What significance should we attach to the agreement between the absorption 
cross sections  of D-branes and of black holes? This
agreement has been demonstrated at low energies and for black holes near the 
extremal limit in \dasmathurtwo\dasmathurthree . 
 (The result of \maldastrom\ extends this to situations 
further from extremality.)  It
is possible that what we are seeing here is  a universal structure of
 low energy 
amplitudes  in a supergravity theory, perhaps for states not too
 far from extremality.

 For example the
 low velocity  scattering of BPS
monopoles is given by geodesics on moduli space, and moduli space is a purely 
BPS construct. Thus the physics of slightly non-BPS processes (that of slowly
 moving monopoles) is capable of being described by knowledge of only BPS 
structures. It is not clear if low energy massless quanta are  similar to
 slow moving massive BPS states, in the sense that moduli space physics 
may capture their interactions.

Recently it has been shown that there is an interesting algebraic 
relation giving the three point couplings of BPS states in string 
theory \ref\harmoore{J.A. Harvey and G. Moore, hepth 9610182,
 hepth 9609017.}. It would be interesting if this could be 
connected to the processes involving the absorption of low energy 
massless particles by massive BPS states.

Most intersting of course is the possibility that the  agreement
 described in section 3 extends to higher energy quanta,  smaller
 compactification radii,   and to black
 holes far from extremality.  An important question in this regard would
 be to understand the structure of BPS bound states
 involving different branes, and the excitations 
around such bound states, as the coupling is taken
 from weak towards strong. These issues are under investigation.

\bigskip
\bigskip
\bigskip

{\bf {Acknowledgements}}
\bigskip

I would like to thank the organisers of this conference
for the opportunity to attend the conference and to give
this talk.
I would like to thank Sumit Das for collaboration on the work
relating to D-branes and  1+1 dimensional string theory, for many
useful discussions and for reviewing this manuscript. I would like 
to thank Esko Keski-Vakkuri for
collaboration on the work related to the semiclassical 
approximation to quantum gravity. I would like to thank  D. Kabat,
 G. Lifschytz,  J. Maldacena, A. Sen,  L Susskind and  B. Zwiebach 
for helpful discussions.
This work is supported in part by D.O.E. cooperative agreement
DE-FC02-94ER40818.

\listrefs
\end